\documentclass[preprint,showpacs,preprintnumbers,amsmath,amssymb]{revtex4}

\usepackage{graphicx}
\usepackage{dcolumn}
\usepackage{bm}

\begin{document}

\preprint{APS/123-QED}

\title{Analytical Solution to Transport in
         Brownian Ratchets via Gambler's Ruin Model}

\author{X. Z. Cheng}
\author{M. B. A. Jalil}
\affiliation{%
Department of Electrical and Computer Engineering, National
University of Singapore, Singapore
}%

\author{Hwee Kuan Lee}
\affiliation{%
Bioinformatics Institute, 30 Biopolis Street, \#07-01 Matrix,
138671, Singapore
}%

\date{\today}

\begin{abstract}

We present an analogy between the classic Gambler's Ruin problem and
the thermally-activated dynamics in periodic Brownian ratchets. By
considering each periodic unit of the ratchet as a site chain, we
calculated the transition probabilities and mean first passage time
for transitions between energy minima of adjacent units. We consider
the specific case of Brownian ratchets driven by Markov dichotomous
noise. The explicit solution for the current is derived for any
arbitrary temperature, and is verified numerically by Langevin
simulations. The conditions for vanishing current and current
reversal in the ratchet are obtained and discussed.

\end{abstract}

\maketitle

In the \textit{Gambler's Ruin }problem, a player plays a series of
games against an adversary, winning (or losing) one dollar for every
success (or failure), until one of them is ``ruined''. Given the
probability of winning each game, the Gambler's Ruin problem
considers the probability of ultimate ruin of one of the players, as
well as the number of games required \cite{feller}. In this paper,
we show an intimate relationship between this classic random walk
problem and the thermally activated dynamics in arbitrary
potentials. The linkage between these two disparate topics is made
possible by recent advances in the time quantification (TQ) of Monte
Carlo (MC) \cite{chengprl,kikuchi}. In particular, the evolutionary
techniques for the Gambler's Ruin problem can be utilized to analyze
the transition probabilities and the mean first passage time (MFPT)
of the complex stochastic transport in Brownian ratchets.

An oscillating driving force applied on Brownian particles in
asymmetric periodic potentials (ratchets) can cause directed
transport, i.e. imbalanced current
\cite{reimann,astumian94,doering}. The keen scientific interest in
the transport property of Brownian ratchets is attributed to their
role in biological systems, e.g. the astonishing energy-motion
conversion of ATP hydrolysis \cite{svoboda}.
%
%
One of the key questions in the study of Brownian ratchets is
obtaining the expression for current.
%
%
In general, the stochastic transport in the ratchets is
modeled by Langevin equations of the form:
\begin{equation}\label{eq:sde}
    \gamma\dot{x} = -U'(x, z(t)) + \xi(t),
\end{equation}
where $\xi(t)$ is a mean zero Gaussian white noise term, i.e.
$\left<\xi(t)\xi(s)\right> = 2\gamma k_BT\cdot \delta(t-s)$, and
$z(t)$ is a Markov dichotomous process with correlation time
$\tau_c$. $\xi(t)$ represents the effects of thermal fluctuation,
while $z(t)$ models stochastic processes such as impurities or
defects jumping between metastable states \cite{kula}. The current
is calculated by solving the corresponding Fokker-Planck equation
(FPE) under periodic boundary conditions. However the explicit
current expression can only be obtained for a few simple cases
\cite{kula,broeck,reimann96}, due to the complexity of dichotomous
processes induced dynamics. For non-trivial cases, the ratchet
current can be calculated by simulating the Langevin equation
\cite{lindner} or from numerical solutions of the FPE \cite{bier96}.

Numerical calculations are computationally intensive and do not
yield as much physical insights as analytical solutions. Our
objective is thus to derive the analytical expression of the current
for an arbitrary ratchet potential. This is performed by first
discretizing each periodic unit of the ratchet into a finite site
chain with absorbing boundaries and analyzing the random walk within
this chain. This is essentially the classic Gambler's Ruin problem,
with some modifications to account for the dichotomous process. To
complete the analogy, the corresponding time (in seconds) of one
Monte Carlo step (MCS) has to be justified. Drawing from recent
advances in quantifying the MCS \cite{chengprl,kikuchi}, e.g. by
linking the MC scheme to the Langevin dynamics via the FPE
\cite{chengprl}, we formulate a time-quantification technique based
on the Central Limit Theorem (CLT). With this, we established the
theoretical basis for the analytical expression of the ratchet
current obtained through the MC approach.

To explain our formalism, we shall first introduce time
quantification of the MCS based on the CLT. Next, we model the
transport in a Brownian ratchets with MC formalism and derive the
transition probability and the MFPT in each periodic unit, and hence
the ratchet current. The current expression is obtained for the
simple thermal equilibrium case, and the more complex case of
dichotomous noise. Finally, by applying the aforementioned TQ factor
the current expression is verified by means of numerical Langevin
simulation.

Time quantification of the MCS is most easily introduced by
considering an overdamped Brownian particle in a steady potential
$U(x,z(t)) = V(x)$. The random walk on $x$ takes a fixed length
trial move: $\Delta x = -R, R$ ($R \ll 1$) with equal trial
probability in both directions but subject to the heat-bath
acceptance rate of $1/(1+\exp(\beta\Delta V))$. Here $\Delta V$ is
the energy difference in the proposed transition and
$\beta\equiv1/k_BT$. Expanding the heat-bath acceptance rate, we
obtain the mean $\mu$ and variance $\sigma$ of $\Delta x$ in
\emph{one} MCS:
$ \mu =\frac{1}{4} \beta f(x_0) R^2$ and $\sigma ^2 = \frac{1}{2}R^2
+ O(R^4)$
, where $f(x) = -V'(x)$ is the external force. Since $R \ll 1$, the
change of $f(x)$ within a few MCS is negligible. According to the
CLT, after a large number $n$ MCS the spread of displacement from
$x_0$ approximates the normal distribution:
\begin{equation}\label{eq:nMCint}
    P(\Delta x_{\mathrm{MC}}) = N(n\mu, n\sigma^2)
                              = f(x_0) \cdot n \frac{1}{4} \beta R^2 + \eta \sqrt{2n\frac{1}{4}R^2},
\end{equation}
where $\eta \!\sim\! N(0,1)$ follows the standard Gaussian
distribution. We note that the integration form (Ito's
interpretation) of the overdamped Langevin equation of Eq.
(\ref{eq:sde}) also results in a normal distribution of the
displacement $\Delta x$ after a time interval $\Delta t$:
\begin{eqnarray}
    P(\Delta x_{\mathrm{LD}}) 
                              &=& \frac{1}{\gamma}f(x_0)\Delta t + \eta \sqrt{2(k_BT/\gamma)\Delta t}.
    \label{eq:sdeint}
\end{eqnarray}
Comparing Eq.~(\ref{eq:nMCint}) and Eq.~(\ref{eq:sdeint}), we obtain
a term-by-term equivalence between $\Delta x_{\mathrm{MC}}$ and
$\Delta x_{\mathrm{LD}}$ if
\begin{equation}\label{eq:TQfactor}
    1\; \mathrm{MCS} = \Delta t /n = \gamma\beta R^2/4.
\end{equation}

Since the dichotomous process $z(t)$ simply produces transitions
between the two potential profiles, the equivalence established in
Eq.~(\ref{eq:TQfactor}) is still valid in the presence of $z(t)$,
subject to the condition that $\Delta t \ll \tau_c$.
%
%
This equivalence justifies the use of MC methods to analyze the
ratchet current instead of the Langevin equation.

\begin{figure}
  \centering
  \includegraphics[width=0.4\textwidth, bb={10 12 210 156}] {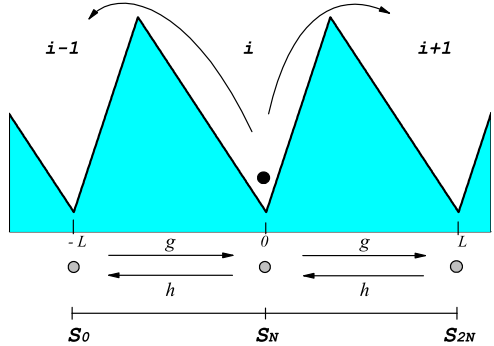}
  \caption{\label{fig:ratchet} Schematic diagram of a $L$-periodic ratchet potential.}
\end{figure}

We consider an $L$-periodic ratchet which can be divided into
``supersites'' of length $L$, e.g. sites $i-1$, $i$ and $i+1$, as
shown in Fig.~(\ref{fig:ratchet}). The forward and backward
transition probability between the supersites, defined as $g\equiv
p(x \to x+L)$ and $h \equiv p(x \to x-L)$, respectively, should be
constant for all $x$ corresponding to an energy minimum, due to the
periodic property. The difference between the transition
probabilities of $g$ and $h$ results in a non-zero current. We hence
have the steady current:
\begin{equation}\label{eq:currentdef}
 \left<\dot{x}\right> := \lim_{t\to\infty} \frac{x(t) - x(0)}{t}
                       = \frac{(g-h)L}{\tau_{\mathrm{\scriptscriptstyle MFPT}}},
\end{equation}
where $\tau_{\mathrm{\scriptscriptstyle MFPT}}$ is the MFPT for the
particle starting at position $x_0$ to hit either position $x_0+L$
or $x_0-L$. $\tau_{\mathrm{\scriptscriptstyle MFPT}}$ is a critical
factor in influencing the transport in the ratchets and has been
studied for several limited cases \cite{reimann96,iwaniszewski}.

We start our analysis with a simple illustrative case -- a thermal
equilibrium Brownian ratchets without a driven noise. We first
discretize the ratchets of length $2L$ into $2N+1$ micro-sites, i.e.
$\{S_0,\ldots,S_{2N}\}$ as illustrated in Fig.~(\ref{fig:ratchet}).
A particle starts at site $S_m$, and moves to adjacent micro-sites
randomly, e.g. with steady probability $\mu_m$ to site $S_{m-1}$ and
with probability $w_m$ to site $S_{m+1}$. We define $g(m)$ as the
probability that the particle from site $S_m$ reach the end site
$S_{2N}$ before it reaches the site $S_0$. We further define
$\tau(m)$ as the MFPT (in MCS) for the particle to reach either end
site $S_0$ or $S_{2N}$. Hence, we obtain the difference relations
for $g(m)$ and $\tau(m)$:
\begin{eqnarray}
    g(m) &=& \mu_m \cdot g(m-1) + w_m \cdot g(m+1) + (1 - \mu_m - w_m)\cdot g(m)
  \label{eq:zerodiffg}\\
    \tau(m) &=& \mu_m \cdot \tau(m-1) + w_m \cdot \tau(m+1) + (1 - \mu_m - w_m)\cdot \tau(m) + 1,
  \label{eq:zerodifftau}
\end{eqnarray}
which are analagous to those of the Gambler's Ruin problem. The
initial conditions $g(0) = 0$, $g(2N) = 1$ and $\tau(0) = \tau(2N) =
0$ apply.

The solution to Eq.~(\ref{eq:zerodiffg}) can be obtained by the
recurrence relation: $g(m+1) - g(m) = (\mu_m/w_m)(g(m) - g(m-1))$.
Starting from the middle minima i.e. $m = N$, we can obtain the
forward transition probability $g$ between the adjacent supersites:
\begin{equation}\label{eq:gsimple}
    g = g(N) = \frac{\sum^{N-1}_{i=0} k(i)}{\sum^{2N-1}_{i=0} k(i)}
             = \frac{1}{1+ k(N)},
\end{equation}
where $k(0)\equiv1$, $k(m) \equiv \prod^{m}_{i=1} \mu_i/w_i$ for $m
\ge 1$. In the last step we have used the periodic condition: $\mu_j
= \mu_{N + j}$ and $w_j = w_{N+j}$, which leads to $k(N+i) = k(N)
\cdot k(i)$. The backward transition probability can be obtained
immediately from $h = (1-g)$. Similarly for
Eq.~(\ref{eq:zerodifftau}), after some simplification we obtain, for
$m = N$:
\begin{equation}
    \tau_{\mathrm{\scriptscriptstyle MFPT}} = \tau(N)
        = g \cdot \sum_{i=1}^{N} \left((w_i k(i))^{-1} \sum_{j=i}^{N+i-1}k(j)\right).
\end{equation}

Substituting the heat-bath rate definition
$\frac{1}{1+\exp(\beta\Delta V)}$ for $w_i$ and $\mu_i$ into $k(i)$,
we obtain: $k(i) = (w_0/w_i)e^{\beta V_i} = 2w_0(e^{\beta V_i} +
e^{\beta V_{i+1}})$, where $V_i$ is the potential at the $i$th site
and $V_0 \equiv 0$. Particularly, $k(N) = \exp(\beta V_N)$ since
$w_0 = w_N$. Thus, by considering Eqs.~(\ref{eq:TQfactor}) and
(\ref{eq:currentdef}), the current expression for ratchets in
thermal equilibrium converges to the well-discussed continuous form
\cite{risken} as $N \to \infty$:
\begin{equation}\label{eq:equilibriumcurrent}
    \left<\dot{x}\right> = \frac{(g-h)L}{\tau_{\mathrm{\scriptscriptstyle MFPT}}}
                         = \frac{L \cdot (1-e^{\beta V(L)})} {\gamma\beta \int_0^L dx \; e^{-\beta V(x)}\int_x^{x+L}dy\; e^{\beta V(y)}}.
\end{equation}

We shall also point out that for $N = 2$, our above discussion
reduces to the three-state discrete-time minimal Brownian ratchet
model \cite{lee}. In particular, if we replace the transition rates
with $\mu'_i = \tilde{\gamma}\mu_i + \gamma/2$ and $w'_i =
\tilde{\gamma} w_i + \gamma/2$ following the definitions in
Ref.~\cite{lee}, we achieve the same current expression via
Eq.~(\ref{eq:currentdef}) directly.

We now extend our discussion to the non-equilibrium case, i.e. with
an additional dichotomous noise $z(t)$ applied to the ratchets
potential. We consider a mean-zero $z(t)$, which takes two discrete
values $\{1, -\theta\} (\theta>0)$ with correlation
$\left<z(t)z(s)\right> = (Q/\tau_c)\exp(-|t-s|/\tau_c)$, where
$Q=\theta\tau_c$. For clarity, we denote ``$+$'' and ``$-$'' as
representing the two states $z = 1$ and $z = -\theta$ respectively.
Similar to our previous analysis, we define $g(m; \sigma; \sigma')$
as the probability for a particle at initial site $S_m$ with $z(t_0)
= \sigma$ to reach the end site $S_{2N}$ after some time $t$ with
$z(t) = \sigma'$ before it reaches site $S_0$. We also define
$\tau(m;\sigma)$ as the MFPT for the particle starting at $S_m$
under $z(t_0)=\sigma$ to reach any end sites.

We first calculate the $g(m;\sigma;\sigma')$, by considering the
following four difference equations:
\begin{equation}\label{eq:dichdiff}
g(m; \sigma; \sigma')= \sum_{\tilde{\sigma}}v(\tilde{\sigma}|\sigma)
                                    \cdot \left[ w_m^{\tilde{\sigma}} g(m + 1; \tilde{\sigma}; \sigma')
                                                 + \mu_m^{\tilde{\sigma}}  g(m - 1; \tilde{\sigma}; \sigma')
                                                 + (1 - w_m^{\tilde{\sigma}} - \mu_m^{\tilde{\sigma}} ) g(m; \tilde{\sigma}; \sigma')
                                          \right],
\end{equation}
where $v(\tilde{\sigma}|\sigma)$ is the transition probability for
dichotomous state from $\sigma$ to $\tilde{\sigma}$ in \emph{one}
MCS \cite{barik}. $w_m^{\sigma}$ and $\mu_m^{\sigma}$ denote the
spatial transition rates at dichotomous state $z(t) = \sigma$.
Equation~(\ref{eq:dichdiff}) can be rewritten into a $2\times2$
matrix difference equation. After some algebra, we obtain:
\begin{equation}\label{eq:mat_diffX}
    X_{m+1} = W_m^{-1} \left( \lambda \cdot C + W_m + U_m \right) \cdot X_{m} - W_m^{-1} U_m \cdot X_{m-1},
\end{equation}
where $\lambda\equiv \frac{v(-|+)}{1-v(-|+) - v(+|-)} \ll 1$, $W_m =
\mathrm{Diag}\{w_m^+, w_m^-\}$, $U_m = \mathrm{Diag}\{\mu_m^+,
\mu_m^-\}$, and
%
\[
C = \left( {\begin{array}{cc}
   {1} & {-1}  \\
   {-\theta} & {\theta}  \\
\end{array}} \right)
; \quad
X_m = \left( {\begin{array}{*{20}c}
   {g(m;+;+) } & {g(m;+;-)}  \\
   {g(m;-;+) } & {g(m;-;-)}  \\
\end{array}} \right).
\]
%
The initial conditions are $X_0 = \mathbf{0}$;
$X_{2N}=\mathrm{Diag}\{1,1\}\equiv I$. The additional correlation
term $(\lambda\cdot C)$ in Eq.~(\ref{eq:mat_diffX}) prevents a
general solution for $X_m$. Nevertheless $X_m$ can be reduced into a
linear combination of $X_1$ and $X_0$ by recurring
Eq.~(\ref{eq:mat_diffX}). Since $X_0 = \mathbf{0}$, we can thus
define a $2 \times 2$ matrix $Q_m$ such that
\begin{equation}
    X_m = Q_m \cdot X_1 \hspace{0.1in} (m\ge 1).
\end{equation}
$Q_m$ can be expressed as a polynomial function of $\lambda$:
\begin{equation}\label{eq:Qm}
    Q_m(\lambda) = \sum_{l=0}^{m-1} D_l^m \cdot \lambda^l,
\end{equation}
where $D_l^m$ are $2\times2$ matrices which are given by:
\begin{eqnarray}
    D_0^m &=& \sum_{i=0}^{m-1} K_i,
\nonumber\\
    D_l^m &=& \sum_{i=l}^{m-1} \left[
        \left( \sum_{j=i}^{m-1} K_j\right)  K_i^{-1} W_i^{-1} C \cdot D_{l-1}^i
    \right],
\end{eqnarray}
in which $K_i = \mathrm{Diag}\{k_i^+, k_i^-\}$ where $k_i^\sigma
\equiv \prod^{i}_{j=1} \mu_j^\sigma/w_j^\sigma$. In the present
application, we are typically interested in the first few factors
since
$\lambda \cong \frac{\Delta t}{(1+\theta)\tau_c} \ll 1 $ \cite{barik}.
Our experience shows that truncating the polynomial at
$O(\lambda^3)$ could already yield a very good approximation to
$Q_m$.

Setting the starting position $m=N$, we obtain the \emph{forward
transition probability matrix}:
\begin{equation}\label{eq:G}
    G = X_N = Q_N \cdot X_1 = Q_N \cdot Q_{2N}^{-1},
\end{equation}
where in the last step, $X_{2N} = Q_{2N}\cdot X_1$ is noted. The
\emph{backward transition probability matrix} $H$ can be calculated
similarly from Eq.~(\ref{eq:mat_diffX}) with the reverse initial
conditions: $X_0 = I$ and $X_{2N} = \textbf{0}$ .

Similarly, the calculation of $\tau(m;\sigma)$ leads to the matrix difference equation:
\begin{equation}\label{eq:mat_diffY}
    W_m\cdot Y_{m+1} = \left(\lambda\cdot C + W_m + U_m\right) \cdot Y_{m} - U_m \cdot Y_{m-1} - E,
\end{equation}
where $Y_m = (\tau(m;+), \tau(m;-))^T$, $E = (1, 1)^T$ and $Y_0 =
Y_{2N} = 0$. Defining $Y_m = Q_m\cdot Y_1 + R_m$, we obtain after
some algebra the \emph{MFPT matrix} $T$ as:
\begin{equation}\label{eq:T}
    T = Y_N =  \left(G\cdot R_{2N}\right) - R_N,
\end{equation}
where $R_m$ is also a polynomial function of $\lambda$ similar to
$Q_m$. The details on $R_m(\lambda)$ will be given elsewhere
\cite{chenginprepare}.

We have now evaluated $G$, $H$ and $T$. However, to calculate the
steady current, we require the effective transition probabilities
$g_{\mathrm{eff}}$ and $h_{\mathrm{eff}}$. We note $S = G + H$ is
the actual \emph{transition matrix for the probability distribution
of dichotomous state} over \emph{one} transition between adjacent
supersites. Hence, the steady state (after $n \to \infty$
transitions) yields the following probabilities of the dichotomous
states at the start of the $(n+1)^{\rm th}$ transition:
$\mathrm{Prob}(z=1) = S_{21}/(S_{12}+S_{21})$ and
$\mathrm{Prob}(z=-\theta) = S_{12}/(S_{12}+S_{21})$. The effective
forward transition probability is then given by $g_{\mathrm{eff}} =
\sum_{\sigma,\sigma'} \mathrm{Prob}(z=\sigma)g(N;\sigma;\sigma')$,
and similarly for $h_{\mathrm{eff}}$. Based on
Eq.~(\ref{eq:currentdef}), this then leads to our main result, i.e.
the analytical expression of the ratchets current:
\begin{equation}\label{eq:currentEFF}
 \left< \dot{x} \right> =  \frac {G_{11} + G_{22} - H_{11} - H_{22} - 2 (|G|-|H|)}
                            {(G_{21}+H_{21})T_1 + (G_{12}+H_{12})T_2}.
\end{equation}

\begin{figure}
\centering
\includegraphics[width=0.44\textwidth, bb={10 10 290 210}]
{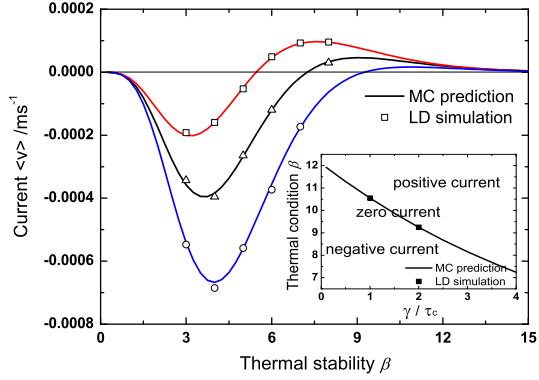}
\caption{\label{fig:current} Temperature-driven reversal of ratchets
current. Close agreement between analytical MC prediction and
Langevin dynamical (LD) simulation. The simulation parameters are:
$R = 0.005$, $L = 1.0$, $F = 0.6$, $\theta = 0.42$, $\gamma=1$ and
$\tau_c = 0.15, 0.25, 0.5$ from top to bottom. Error bars are
smaller than the symbol size. Inset: extracted zero-current curve
with respect to $\gamma / \tau_c$.}
\end{figure}

For verification, we performed a numerical simulation based on the
Langevin equation of Eq. (\ref{eq:sde}), and assuming a ratchet
potential profile of:
\begin{equation}
    U(x,z(t)) =
    \left\{\begin{array}{ll}
        -\frac{1}{\hat{k}} L\hat{x}  - z(t)\cdot Fx;\quad &\hat{x} \le \hat{k} \\
        \frac{1}{1-\hat{k}} L \hat{x} - z(t)\cdot Fx;\quad &\hat{x} > \hat{k}
    \end{array}\right.,
\end{equation}
where $\hat{x} = x - [x/L]$, and $\hat{k} = 2/3$ reflects the
asymmetry of the potential \cite{kula}.
%
%
In Fig.~(\ref{fig:current}), we plotted the particle current from
the Langevin simulation and found extremely close agreement with the
predictions of Eq.~(\ref{eq:currentEFF}).

\begin{figure}
\centering
\includegraphics[width=0.44\textwidth, bb={55 30 280 215}]
{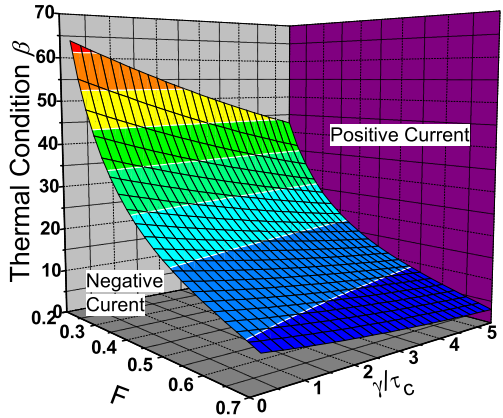}
\caption{\label{fig:zerocurve} The zero-current surface with respect
to parameters $\beta$, $\gamma/\tau_c$ and $F$.}
\end{figure}

Asymmetry in potential profile and dichotomous fluctuations can
result in current reversal \cite{kula}. The MC method enables us to
obtain precisely the vanishing current condition (see the inset of
Fig.~\ref{fig:current}) which is of importance in rectifying
particles with only small differences in $\gamma$. Interestingly,
since $\lambda \cong \frac{\Delta t}{(1+\theta)\tau_c} =
\frac{\gamma}{\tau_c}\frac{\beta R^2}{4(1+\theta)}$, from
Eq.~(\ref{eq:Qm}) we found $(\gamma/\tau_c)$ determines the current
direction.
In Figs.~(\ref{fig:current}) and (\ref{fig:zerocurve}), we observed
two facts: 1) there is a \emph{threshold temperature} $\beta_c$,
below which no current reversal can occur regardless of $\gamma$ and
$\tau_c$, and 2) the zero-current condition curve is
\emph{monotonic} in character, i.e. a decrease in the required
$\gamma / \tau_c$ with increasing $\beta$. These can be explained by
considering the energy barrier between the supersites ($\Delta
V^\sigma_b$) induced by the dichotomous noise. In the present
application, $\Delta V^+_b < \Delta V^-_b$, and hence a positive current
occurs in the limit of high $\beta$. While at \emph{low} $\beta$ and
\emph{large} $\tau_c$ limit that $\tau_{\mathrm{\scriptscriptstyle
MFPT}} \ll \tau_c$, a negative current will be formed if
[$\exp(-\beta \Delta V^+_b)/\theta]
< \exp(-\beta \Delta V^-_b)$. Therefore, the bottom-left (top-right)
corner of the phase diagram of Fig.~(\ref{fig:zerocurve})
corresponds to a negative (positive) current region, thus implying a
monotonic trend of the zero-current surface dividing the two
regions.

Note that the analytical ratchet current in
Eq.~(\ref{eq:currentEFF}) is derived without the assumption of low
temperature as in \cite{reimann96}. Additionally, the above MC
method can reasonably be extended to ratchets driven by an $n$-state
process or even an Ornstein-Uhlenbeck (O-U) process \cite{risken}
(an O-U process is equivalent to an infinite $n$-state process from
the MC point of view). With some modifications, the MC analysis can
also be applied to model the temperature (generalized
Smoluchowski-Feynman) ratchets \cite{reimann}.

To summarize, we presented a time-quantified MC method, based on and
extended from the Gambler's Ruin problem, to analyze the directed
transport in overdamped Brownian ratchets. By considering the
transition probabilities and the MFPT between the adjacent minima of
the periodic ratchet, we derived the analytical expression for the
current in the presence of dischotomous noise, as well as the
vanishing current condition.  Generally, the MC formalism offers an
alternative way to solve intractable stochastic dynamics and the
corresponding Fokker-Planck equations. Extensions to the classic
Gambler's Ruin or other MC problems, e.g. inclusion of correlations
\cite{bohm} or multiple currencies \cite{orr}, may yield further
insights into other areas of stochastic dynamics, e.g. turbulence or
high-dimensional thermally activated dynamics.


\end{document}